%% file: main.tex
\documentclass[preprint,12pt]{elsarticle}

\usepackage{amssymb}
\usepackage{amsmath,amsfonts}
\usepackage{graphicx}
\usepackage{textcomp}
\usepackage{booktabs}
\usepackage{multirow}
\usepackage{tabularx}
\usepackage{xcolor}
\definecolor{green}{rgb}{0.09, 0.45, 0.27}
\usepackage{comment}
\usepackage{url}
\usepackage{booktabs}

\usepackage{todonotes}

\newlength\MAX  
\newcommand*\Chart[1]{#1~\rlap{\textcolor{black!20}{\rule{\MAX}{2ex}}}\rule{#1\MAX}{2ex}}

\usepackage{hyperref}

\journal{Expert Systems with Applications}

\begin{document}

\begin{frontmatter}

%% Title, authors and addresses

%% use the tnoteref command within \title for footnotes;
%% use the tnotetext command for theassociated footnote;
%% use the fnref command within \author or \address for footnotes;
%% use the fntext command for theassociated footnote;
%% use the corref command within \author for corresponding author footnotes;
%% use the cortext command for theassociated footnote;
% use the ead command for the email address,
% and the form \ead[url] for the home page:
% \title{Title\tnoteref{label1}}
% \tnotetext[label1]{}
% \author{Name\corref{cor1}\fnref{label2}}
% \ead{email address}
% \ead[url]{home page}
% \fntext[label2]{}
% \cortext[cor1]{}
% \affiliation{organization={},
%             addressline={},
%             city={},
%             postcode={},
%             state={},
%             country={}}
% \fntext[label3]{}

\title{Who Evaluates the Evaluators? On Automatic Metrics for Assessing AI-based Offensive Code Generators}

%% use optional labels to link authors explicitly to addresses:
%% \author[label1,label2]{}
%% \affiliation[label1]{organization={},
%%             addressline={},
%%             city={},
%%             postcode={},
%%             state={},
%%             country={}}
%%
%% \affiliation[label2]{organization={},
%%             addressline={},
%%             city={},
%%             postcode={},
%%             state={},
%%             country={}}

\author[label1]{Pietro Liguori}
\ead{pietro.liguori@unina.it}
\author[label1]{Cristina Improta\corref{cor}}
\ead{cristina.improta@unina.it}
\author[label1]{Roberto Natella}
\ead{roberto.natella@unina.it}
\author[label2]{Bojan Cukic}
\ead{bcukic@uncc.edu}
\author[label1]{Domenico Cotroneo}
\ead{cotroneo@unina.it}

\address[label1]{University of Naples Federico II, Naples, Italy}
\address[label2]{University of North Carolina at Charlotte, North Carolina, US}
%\affiliation[label1]{organization={University of Naples Federico II},
            %addressline={},
             %city={Naples},
             %postcode={},
             %state={},
             %country={Italy}}
%\affiliation[label2]{organization={University of North Carolina at Charlotte},
            %city={Charlotte},
            %state={North Carolina},
            %country={US}} 

\cortext[cor]{Corresponding author}

\begin{abstract}
AI-based code generators are an emerging solution for automatically writing programs starting from descriptions in natural language, by using deep neural networks (Neural Machine Translation, NMT). In particular, code generators have been used for ethical hacking and offensive security testing by generating proof-of-concept attacks.
Unfortunately, the evaluation of code generators still faces several issues. The current practice uses output similarity metrics, i.e., automatic metrics that compute the textual similarity of generated code with ground-truth references. However, it is not clear what metric to use, and which metric is most suitable for specific contexts.

This work analyzes a large set of output similarity metrics on offensive code generators. We apply the metrics on two state-of-the-art NMT models using two datasets containing offensive assembly and Python code with their descriptions in the English language. We compare the estimates from the automatic metrics with human evaluation and provide practical insights into their strengths and limitations.

\end{abstract}

%%Graphical abstract
%\begin{graphicalabstract}
%\includegraphics{grabs}
%\end{graphicalabstract}

%%Research highlights
\begin{comment}
\begin{highlights}
\item There is a need for automated metrics to evaluate neural machine translation models
\item We assess the accuracy of automated metrics for offensive security code
\item N-gram-based metrics become less accurate as N decreases
\item The ``exact match'' metric is the most accurate for assembly code
\item The ``edit distance'' metric is the most accurate for Python code

\end{highlights}
\end{comment}

\begin{keyword}
AI-based Code Generators \sep Offensive Code \sep Neural Machine Translation \sep Software Security \sep Output Similarity Metrics

%% PACS codes here, in the form: \PACS code \sep code

%% MSC codes here, in the form: \MSC code \sep code
%% or \MSC[2008] code \sep code (2000 is the default)

\end{keyword}

\end{frontmatter}

%% \linenumbers

%% main text
\section{Introduction}
\label{sec:intro}
\input{introduction}

\section{Related Work}
\label{sec:related}
\input{related}

\section{Offensive Code Generation}
\label{sec:code_generation}
\input{code_generation}

\section{Code Generation Metrics}
\label{sec:metrics}
\input{metrics}

\section{Experimental Setup}
\label{sec:case_study}
\input{case_study}

\section{Experimental Results}
\label{sec:experiments}
\input{experiments}

\section{Discussion}
\label{sec:discussion}

\input{discussion.tex}

\section{Threats to Validity}
\label{sec:threats}
\input{threats}

\section{Conclusion}
\label{sec:conclusion}
\input{conclusion}

\section*{Acknowledgements}
This work has been partially supported by the University of Naples Federico II in the frame of the Programme F.R.A., project OSTAGE, PROT: 34938\_07\_04\_2021, UGOV: 000010-ALTRI\_CdA\_75\_2021\_FRA\_LINEA\_B\_001\_002, CUP: E55F21000340005.
%% The Appendices part is started with the command \appendix;
%% appendix sections are then done as normal sections
%% \appendix

%% \section{}
%% \label{}

%% For citations use: 
%%       \citet{<label>} ==> Jones et al. [21]
%%       \citep{<label>} ==> [21]
%%

%% If you have bibdatabase file and want bibtex to generate the
%% bibitems, please use
%%
  \bibliographystyle{elsarticle-harv}
  \biboptions{authoryear}
  \bibliography{mybibfile}

%% else use the following coding to input the bibitems directly in the
%% TeX file.

%% \begin{thebibliography}{00}

%% \bibitem[Author(year)]{label}
%% Text of bibliographic item

%% \bibitem[ ()]{}

%% \end{thebibliography}
\end{document}

%% file: introduction.tex
\textit{Offensive AI}, i.e., the (ab)use of Artificial Intelligence (AI) to accomplish a malicious goal, is an emerging threat for computer systems~\citep{mirsky2022threat}. 
Indeed, AI is paving the way for a new generation of offensive security techniques, by helping adversaries to launch attacks that were not possible before. 
Suffice it to think that AI has been adopted to conduct spear-phishing attacks~\citep{stupp2019fraudsters,mirsky2021creation}, to find zero-day vulnerabilities in software~\citep{mokhov2014use,lin2020software}, to automate reverse engineering~\citep{bao2014byteweight,ding2019asm2vec}, to build realistic fake personas~\citep{salminen2019future,salminen2020enriching}, and many other malicious activities.

AI can also be applied for defensive purposes. 
Indeed, \textit{AI-based security code generators} is an emerging use of AI for supporting security auditors. In general, code generators use machine learning to produce programs (\textit{code snippets}) starting from descriptions (\textit{intents}) in natural language (NL). 
In particular, Neural Machine Translation (NMT) is the state-of-the-art solution using neural networks for code generation~\citep{akinobu2021neural,akinobu2022nmt}. 
In the context of ethical hacking and offensive security testing, AI-based code generators support auditors to develop \textit{proof-of-concept} (POC) attacks, e.g., in order to assess the severity and exploitability of software vulnerabilities and to motivate vendors and users to adopt mitigations~\citep{arce2004shellcode}.

%As in many other areas of artificial intelligence, deep neural networks are bringing impressive improvements also in the translation quality~\citep{koehn2020neural}.
%%\textit{Neural Machine Translation} (NMT) is a powerful approach to machine translation that uses deep neural networks to translate a sequence of words from a source language into a different, target one. 
%%\textit{Code generation}, i.e., the use of NMT models to automatically generate \textit{code snippets} starting from natural language (NL) \textit{intents}, has become attractive to different research areas, from software engineering to software security applications. Examples are program repair~\citep{harer2018learning,jiang2021cure}, coding assistance~\citep{akinobu2021neural,akinobu2022nmt}, generation of software exploits~\citep{liguori2021evil,liguori2022can,YANG2022111577}, etc.  
%As a matter of fact, NMT models finds also application in the context of software security to automatically generate software exploits from natural language (NL) descriptions~\citep{liguori2021evil,liguori2022can,yang2022dualsc}.

An underrated aspect of AI-based code generation is the evaluation of the quality of the output generated by the models~\cite{shterionov2018human}. 
Ideally, users manually evaluate whether the generated code correctly translates the NL intent. Unfortunately, the \textit{human evaluation} is often unfeasible due to the massive amount of data to analyze, which makes the analysis time-consuming and prone to human errors. 
Some studies addressed this problem by introducing a large number of \textit{output similarity} metrics, i.e., automatic metrics computed by comparing the textual similarity of generated code with a ground-truth reference. These metrics are an appealing solution to estimate the quality of generated code since they are reproducible, easily tuned, and time-saving. 
However, they do not fully reflect the correctness of the outputs since generated code can be different from the reference but still correct (e.g., the assembly conditional jumps \texttt{jz} and \texttt{je} are different instructions that can be used to perform the same operation). 
Furthermore, there is no clear indication whether there is a unique metric suitable for any evaluation, or whether a specific metric should be selected depending on the context, such as the programming language of generated code and its application domains (e.g., code generators for ethical hacking). As a result, these output similarity metrics have been used inconsistently to evaluate code generators, making it difficult to compare the performance of different ML models.

Since the choice of the right metric may be more important than the choice of the models used to solve the task~\citep{jiang2008comparing}, it is necessary to understand what metrics should be used and when. 
This work provides a practical assessment of the output similarity metrics commonly used to evaluate NMT models for code generation. The key idea is to compare the output similarity metrics with the human evaluation, in order to discuss the strengths and limitations of the metrics and to identify the most suitable ones for different contexts.
%

%Given the increasing interest for the NMT models in software security~\citep{liguori2021evil,liguori2022can,yang2022dualsc}, we contextualize this work in the automatic generation of software exploits.
%To this aim, we used two datasets in the software security field used to generate offensive assembly and Python code snippets, respectively, starting from the description of the code in the English language. 
In this work, we present an extensive analysis of automatic metrics for evaluating security-oriented code generators. We study $23$ automatic metrics from the literature, by applying them to evaluate two state-of-the-art NMT models. We train and test the NMT models using two datasets of offensive assembly and Python code annotated with descriptions in the English natural language.
Then, we estimate the performance of NMT models using the automatic metrics and compare the results with results from human evaluation as a reference. In summary, this work provides the following key contributions:

\begin{enumerate}
    \item A detailed study of the output similarity metrics most commonly used to evaluate AI-based code generators;
    \item A systematic evaluation, including both quantitative analysis and correlation analysis, of the output similarity metrics, by comparing them with human evaluation;  
    \item Practical insights on what metric to use and when to properly assess the generation of offensive code. We summarize our main findings and their implications in \tablename{}~\ref{tab:findings}.
\end{enumerate}

In the following, Section~\ref{sec:related} discusses the related work; 
Section~\ref{sec:code_generation} describes the code generation task;
Section~\ref{sec:metrics} shows the metrics used to evaluate the NMT models in the code generation; 
Section~\ref{sec:case_study} presents the case study;
Section~\ref{sec:experiments} shows the experimental results;
Section~\ref{sec:discussion} describes the results of our analysis;
Section~\ref{sec:threats} discusses the threats to validity;
Section~\ref{sec:conclusion} concludes the paper.

\begin{table}[t]
\caption{Main findings.}
\label{tab:findings}
\centering
\footnotesize
\begin{tabular}{
>{\centering\arraybackslash}m{2.75cm} |
>{\centering\arraybackslash}m{9.75cm}} 
\toprule
\textbf{Analysis} & \textbf{Main Findings} \\ \midrule
\textit{Statistics on the output similarity metrics} & The output similarity metrics provide very different results on the same test data, making the interpretation of the model's performance very difficult. The metrics overestimate the performance on the assembly data, while they underestimate the performance of the models on the Python data.\\ \midrule
\textit{Quantitative analysis on the whole test data} & 
Metrics based on n-grams such as \textit{ROUGE} and \textit{BLEU} provide estimates close to human evaluation when the number $n$ is low. 
The difference between automatic metrics and human evaluation always worsens when $n$ increases for Python code.\\ \midrule
\textit{Quantitative analysis on correct and incorrect predictions} & \textit{ROUGE-4} and \textit{BLEU-4}, which are among the most commonly used metrics in NMT applications, provide poor estimates when they are focused only on the evaluation of semantically-correct and semantically-incorrect generated code. \\ \midrule % \textit{Compilation accuracy} and \textit{exact match} metrics are the best metrics when focusing on semantically correct and semantically incorrect generated code, respectively, but provide poor estimates of when their focus is switched (i.e., applied to semantically-incorrect and semantically-correct generated code, respectively). \\ \midrule
\textit{Correlation Analysis} & \textit{Exact match} and \textit{edit distance} are the metrics most correlated to the human evaluation for assembly and Python offensive code, respectively. \textit{ROUGE-4} and \textit{BLEU-4} have the lowest correlation regardless of the programming language.\\
\bottomrule
\end{tabular}
\end{table}

%% file: related.tex
%The assessment of the automatic metrics on machine learning tasks is not a new research problem.
\cite{stent2005evaluating} compared the performance of several automatic evaluation metrics using a corpus of automatically generated paraphrases. They showed that the evaluation metrics can at least partially measure similarity in meaning, but are not good measures for syntactic correctness.
\cite{jiang2008comparing} compared the performance of predictive models using design-level metrics with those that use code-level metrics on different datasets from NASA for software fault prediction. The authors showed that the performance of predictive models varies more as a result of using different software metrics groups than from using different machine learning algorithms. 
\cite{8651396} evaluated 121 existing and new metrics, including code-related, documentation-related, and developer-related metrics. They assessed the correlation between each metric and code understandability. The authors concluded that these metrics, even when combined, are not suited to capture the complexity of code and are not suitable for practical applications. 

More recently, the arousing interest in the NMT to solve different tasks pointed out the need for new metrics that can correlate more closely to human evaluation to properly evaluate the translation quality of the models.
\cite{shterionov2018human} compared the scores of three automatic metrics with the results of the human evaluation. They performed an extensive empirical evaluation of the translation task from English to five different natural languages and showed that the automatic metrics underestimate the translation quality of the NMT. 
Similarly, \cite{shimorina2018human} showed that, in the task of natural language generation, the sentence-level correlation between human and automatic metrics is low. %which in turn suggests the need for new automatic evaluation metrics that would better correlate with human scores at the sentence level.
\cite{rao-tetreault-2018-dear} found that the automatic metrics do not correlate well with human judgments in the \textit{style transfer}, i.e., the task of automatically transforming a piece of text in one particular style into another, and thus should be avoided in system development or final evaluation.
\cite{moramarco-etal-2022-human} assessed the correlation between the automatic metrics and human evaluation in the context of the automatic generation of consultation notes from the verbatim transcript of the consultations. The authors showed that all the metrics display a strong bias toward the choice of reference.
\cite{hu2022correlating} conducted experiments in the automatic generation of code documentation and pointed out the low correlation between automatic metrics and human judgments. %Their study pointed out the need to develop specialized automated evaluation metrics that can correlate more closely to human evaluation metrics.
\cite{DBLP:conf/sigsoft/RoyFA21} worked in the context of \textit{code summarization} (i.e., the task of creating readable summaries describing the functionality of the code) to provide a critical evaluation of the applicability and interpretation of automatic metrics as evaluation techniques. They concluded that more reliable metrics should be adopted as new standard metrics for the evaluation.

In the field of code generation, \cite{DBLP:journals/corr/abs-2208-03133} investigated what metric best correlates to a human evaluation in assessing the quality of code generated by NMT models. To address this problem, they considered $6$ metrics to evaluate multiple models for generating Python code snippets from NL descriptions. %They concluded that popular metrics such as \textit{BLEU} and code-oriented metrics such as \textit{CodeBLEU} are not suited to properly evaluate code. 
\cite{DBLP:conf/profes/TakaichiHMKKKT22} used $4$ different metrics to assess the ability of an NMT model to generate Python code from software requirements written in English. %They concluded that \textit{METEOR} best correlates with human judgment on code.
%\todo[noline,size=\footnotesize]{\textbf{Rev 1.2}}\revision{
Other work on code generation resorted to \textit{functional correctness} to evaluate the quality of the generated programs, where a code sample is considered correct if it passes a set of unit tests. 
\cite{kulal2019spoc} used an evaluation metric based on functional correctness to address the problem of producing correct code starting from pseudocode. They generated \textit{k} code samples per problem and assessed the ratio of problems in which any of the \textit{k} samples passed the set of unit tests.
\cite{Chen2021EvaluatingLL} proposed \texttt{pass@k}, an unbiased and numerically stable implementation of this metric. They generated $n \geq k$ samples per task ($n = 200$ and $k \leq 100$), counted the number of correct samples $c \leq n$ that pass unit tests, and calculated an unbiased estimator to benchmark their models in the generation of Python programs from docstrings.
To estimate the functional correctness of a program, however, a set of unit tests needs to be manually constructed. This requires a significant effort that is often unfeasible for large amounts of generated code.%}
%\cite{roziere2020unsupervised} also consider functional correctness to assess the quality of code translated from one programming language to another. unsupervised learning

%State-of-the-art is currently dealing with the applicability of automatic metrics for code generation, however, targeting different domains. 
Similar to \cite{DBLP:journals/corr/abs-2208-03133, DBLP:conf/profes/TakaichiHMKKKT22}, we assess the correlation between the automatic and human evaluation in the code generation, but we adopt a more exhaustive set of $23$ output similarity metrics commonly used to assess NMT models.
Moreover, to the best of our knowledge, this is the first work evaluating the automatic metrics in the generation of code for software security applications (in both low-level and high-level programming languages) and provides practical insights on what metric to use and when to assess the AI-based solutions generating offensive code. 
%The datasets used in our experiments provide NL descriptions both at block and statement levels, which are closer to the descriptions needed for more complex programming tasks. 
For the above-stated reasons, our work can be considered complementary to previous studies. 

%Different from previous work, our work addresses the issue of automatic evaluation of code snippets by employing a comprehensive set of $23$ output similarity metrics commonly used to assess the quality of the programming code generated by NMT models. 
%Furthermore, we assess the correlation between these metrics and human evaluation targeting, besides Python, a low-level programming language such as assembly. 
%To the best of our knowledge, this is the first work to encompass a collection of automatic metrics this extensive in the code generation task for software security. Therefore, it can be considered complementary to previous work.

%% file: code_generation.tex
%\todo[noline,size=\footnotesize]{\textbf{Rev 2.1}}\revision{
\cite{yang2023exploitgen} gathered question posts with tags containing terms related to software exploits from Stack Overflow and counted the number of answers to these posts as well as the number of answers accepted by the questioner. The final results show that less than half of the posts contain answers that were accepted by the questioners. This shows that, due to the specific domain of exploit code, writing exploit code manually is a time-consuming and difficult task.
Therefore, the usage of NMT models to improve developers' productivity and support security auditors in the generation of PoC exploits is becoming an attracting solution~\citep{liguori2021evil,yang2022dualsc,yang2023exploitgen}.%} 
These models generate programming \textit{code snippets} starting from NL \textit{intents}.

\begin{figure}[t]
    \centering
    \includegraphics[width=1\columnwidth]{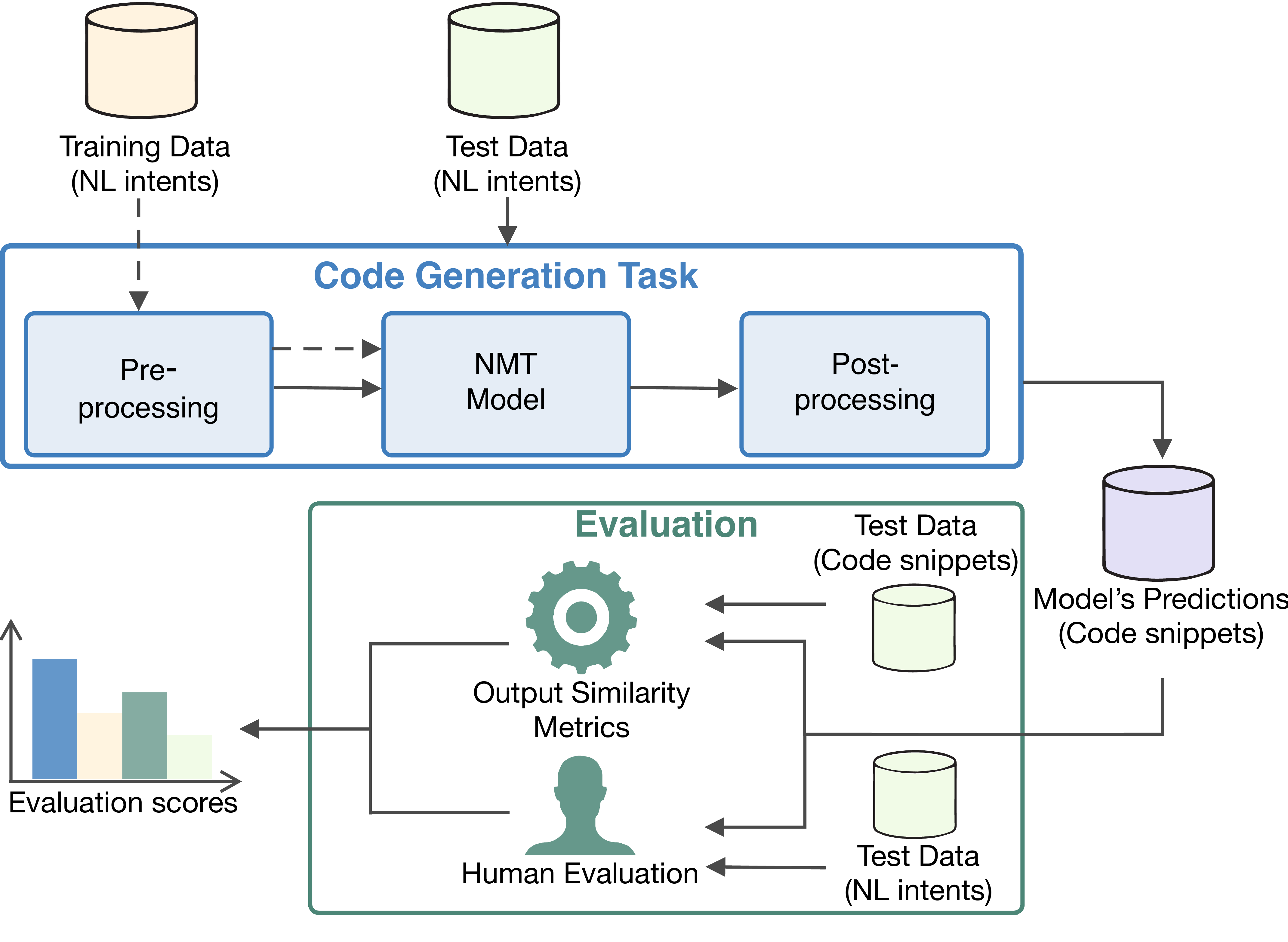}
    \caption{Code generation task.}
    \label{fig:model}
\end{figure}

To perform a rigorous evaluation of the output similarity metrics on the offensive code generated by the models, we follow the best practices in the field.
Hence, we support the models with \textit{data processing} operations. Data processing is an essential step to support the NMT models in the automatic code generation and refers to all the operations performed on the data used to train, validate and test the models.
%To support automatic code generation, neural machine translation is usually accompanied by data processing steps \citep{park2020decoding,8713737,oudah2019impact}. 
These operations strongly depend on the specific source and target languages to translate. 
The data processing steps are usually performed both before translation (\textit{pre-processing}), to train the NMT model and prepare the input data, and after translation (\textit{post-processing}), to improve the quality and the readability of the code in output.
\figurename{}~\ref{fig:model} summarizes the steps.

First, we use a corpus to train the NMT models. The \textit{training data} is pre-processed before being used to feed the model. 
%During the pre-processing, we remove the \textit{stopwords filtering}, i.e., we remove a set of custom compiled words (e.g., the, each, onto) from the intents to include only relevant data for machine translation. Next, we use a tokenizer to break the intents into chunks of information (i.e., the \textit{tokens}) that can be considered discrete elements. Then, we perform the \textit{standardization} of the intents by using an \textit{intent parser} that provides a dictionary of standardizable tokens, such as specific values, label names, and parameters, extracted through regular expressions. The \textit{Standardizer} replaces the selected tokens in the intent with ‘‘\textit{var}\#'', where \# denotes a number from $0$ to \textit{$|l|$}, and $|l|$ is the number of tokens to standardize. The last operation of pre-processing is the generation of the \textit{word embeddings}. 
The pre-processing starts with the \textit{stopwords filtering}, i.e., we remove a set of custom-compiled words (e.g., \textit{the}, \textit{each}, \textit{onto}) from the intents to include only relevant data for machine translation. 
Next, we use a \textit{tokenizer} to break the intents into chunks of text containing space-separated words (i.e., the \textit{tokens}). 
To improve the performance of the machine translation~\citep{li2018named,modrzejewski2020incorporating,liguori2022can}, we \textit{standardize} the intents (i.e., we reduce the randomness of the NL descriptions) by using a \textit{named entity tagger}, which returns a dictionary of \textit{standardizable} tokens, such as specific values, label names, and parameters, extracted through regular expressions. We replace the selected tokens in every intent with ``\textit{var}\#", where \# denotes a number from $0$ to \textit{$|l|$}, and $|l|$ is the number of tokens to standardize.
%Then, we perform the \textit{standardization} of the intents by using an \textit{intent parser} that provides a dictionary of standardizable tokens, such as specific values, label names, and parameters, extracted through regular expressions. 
%The \textit{standardizer} replaces the selected tokens in every intent with ``\textit{var}\#", where \# denotes a number from $0$ to \textit{$|l|$}, and $|l|$ is the number of tokens to standardize. 
Finally, the tokens are represented as real-valued vectors using \textit{word embedding}.  
The pre-processed data is used to feed the NMT model. 
Once the model is trained, we perform the code generation from NL. Therefore, when the model takes as inputs new intents from the \textit{test data} (i.e., data of the corpora not used in the training phase), it generates the related code snippets based on the knowledge inferred during the training (\textit{model's prediction}).
As for the intents, also the code snippets predicted by the models are processed (\textit{post-processing}) to improve the quality and readability of the code. First, the dictionary of standardizable tokens is used in the \textit{de-standardization} process to replace all the ``\textit{var}\#" with the corresponding values, names, and parameters. 
%The output produced by the NMT model is further processed (\textit{post-processing}) to improve the quality and readability of the code. First, the dictionary generated by the intent parser is used in the \textit{de-standardization} process to replace all the ‘‘\textit{var}\#'' with the corresponding values, names, and parameters. Moreover, code snippets are cleaned using regular expressions to remove any extra characters, while the tokens are joined to form complete snippets. 

Finally, the code snippets generated during the model's prediction are evaluated to assess the quality of the code generation task. The evaluation can be performed through output similarity metrics or manual analysis (human evaluation).
The former estimates the quality of the prediction by comparing the model's predictions with the ground truth reference in the test data, the latter, instead, assesses if the output predicted by the model is the correct translation of the NL intent into the generated code snippet.

%% file: metrics.tex
\subsection{Output Similarity Metrics}
\label{subsec:automatic}

Given the huge amount of data to scrutinize, which makes the human evaluation time-consuming, the most practical and common solution to assess the performance of the NMT models is to use metrics that estimate the similarity between the code generated by NMT models and a \textit{ground-truth} (i.e., code snippets used as references for the evaluation). 
%To find the output similarity metrics that best correlate to human judgment in the code generation context, we performed a thorough analysis of the related work of recent years. 
Table~\ref{tab:paper_metrics} presents the most commonly used metrics by previous work to assess the quality of code generated by the NMT models across multiple code-related tasks, including code generation (i.e., natural language to code), code translation (i.e., programming language to different programming language), and code completion (i.e., programming language to the same programming language). In the following, we describe these metrics:
%As the table shows, BLEU is one of the most popular metrics adopted to estimate code correctness, yet its inadequacy in the code-related domain has been pointed out repeatedly \citep{DBLP:conf/sigsoft/RoyFA21, DBLP:conf/profes/TakaichiHMKKKT22,DBLP:journals/corr/abs-2208-03133 }.
%To address this issue, related work employs other automatic metrics, which were designed to explicitly address the weakness in BLEU for the evaluation of natural language sentences (e.g., METEOR, ROUGE), and also for assessing programming language snippets. 
%To provide a comprehensive evaluation, we considered the following output similarity metrics:

\begin{table}[t]
\caption{Output similarity metrics used in previous work to estimate the code generated by NMT models.}
\label{tab:paper_metrics}
\centering
\footnotesize
\begin{tabular}{ 
>{\centering\arraybackslash}m{2.5cm} | 
>{\raggedright\arraybackslash}m{7cm} 
>{\raggedright\arraybackslash}m{3cm}}
\toprule
\centering \textbf{Output Similarity Metric} & \centering \textbf{Work} & \textbf{Code Generation Task} \\ \midrule

\textit{Compilation Accuracy} & \cite{clement-etal-2020-pymt5},  \cite{liguori2021evil} &  NL$\rightarrow$Python; NL$\rightarrow$assembly \\ \midrule

\textit{ROUGE} &
\cite{DBLP:conf/sigsoft/SvyatkovskiyDFS20},  \cite{DBLP:journals/corr/abs-2208-03133}, \cite{clement-etal-2020-pymt5}, \cite{DBLP:conf/profes/TakaichiHMKKKT22} & code completion; NL$\rightarrow$Python\\ \midrule

\textit{BLEU} &
\cite{DBLP:conf/acl/GuoLDW0022},  \cite{DBLP:journals/corr/abs-2105-08645},  \cite{DBLP:conf/naacl/AhmadCRC21},  \cite{DBLP:conf/emnlp/0034WJH21},  \cite{DBLP:journals/corr/abs-2208-03133},  \cite{liguori2021evil}, \cite{clement-etal-2020-pymt5}, \cite{DBLP:conf/profes/TakaichiHMKKKT22} & 
NL$\rightarrow$Java; C\#$\rightarrow$Java; NL$\rightarrow$Python; NL$\rightarrow$assembly \\ \midrule

\textit{Exact Match} &
\cite{DBLP:conf/sigsoft/ChakrabortyADDR22},
\cite{DBLP:conf/acl/GuoLDW0022},  \cite{DBLP:conf/sigsoft/WangYGP0L22},  \cite{DBLP:journals/corr/abs-2105-08645},  \cite{DBLP:conf/naacl/AhmadCRC21}, \cite{DBLP:conf/emnlp/0034WJH21}, \cite{liguori2021evil} &
NL$\rightarrow$Java; NL$\rightarrow$assembly; NL$\rightarrow$Python\\ \midrule

\textit{METEOR} & \cite{DBLP:journals/corr/abs-2208-03133}, \cite{DBLP:conf/profes/TakaichiHMKKKT22} & NL$\rightarrow$Python \\ \midrule

\textit{Edit Distance} & 
\cite{DBLP:conf/acl/GuoLDW0022}, \cite{DBLP:conf/sigsoft/SvyatkovskiyDFS20}, \cite{DBLP:conf/profes/TakaichiHMKKKT22} & code completion; NL$\rightarrow$Python \\ %\midrule

%CodeBLEU &
%Chakraborty \textit{et al.} \cite{DBLP:conf/sigsoft/ChakrabortyADDR22}, Wang \textit{et al.} \cite{DBLP:conf/sigsoft/WangYGP0L22}, Phan \textit{et al.} \cite{DBLP:journals/corr/abs-2105-08645}, Ahmad \textit{et al.} \cite{DBLP:conf/naacl/AhmadCRC21}, Wang \textit{et al.} \cite{DBLP:conf/emnlp/0034WJH21}, Evtikhiev \textit{et al.} \cite{DBLP:journals/corr/abs-2208-03133} &
%NL-Java, C\#-Java, NL-Python \\

\bottomrule
\end{tabular}
\end{table}

\vspace{0.1cm}
\noindent
$\blacksquare$ \textbf{Compilation Accuracy (CA)}. It indicates whether each code snippet produced by the model is compilable according to the syntax rules of the target language. CA value is either $1$, when the snippet's syntax is correct, or $0$ otherwise.

\vspace{0.05cm}
\noindent $\blacksquare$ \textbf{Recall-Oriented Understudy for Gisting Evaluation (ROUGE)}~\citep{lin-2004-rouge}. It measures the matching n-grams (i.e., the adjacent sequence of $n$ items, such as syllables, letters, words, etc.) between the output predicted by the model and the ground truth reference, where $n$ stands for the number of the n-gram. 
The number of matching n-grams is then divided by the total number of n-grams in the reference (recall, \textit{ROUGE R}), or by the total number of n-grams in the model's prediction (precision, \textit{ROUGE P}). The harmonic mean of precision and recall defines $F_1$ Score (\textit{ROUGE $F_1$}).

\vspace{0.05cm}
\noindent $\blacksquare$ \textbf{ROUGE-L}. It is a variant of the ROUGE metric commonly used to assess code generation based on the longest common subsequence (LCS) between the model's output and the reference, i.e. the longest sequence of words (not necessarily consecutive, but still in order) that is shared between both. ROUGE-L recall, precision, and F1-score can be computed by replacing each n-gram match with the LCS.
The ROUGE metrics range between $0$ (perfect mismatch) and $1$ (perfect matching).

\vspace{0.05cm}
\noindent
$\blacksquare$ \textbf{Bilingual Evaluation Understudy (BLEU) score}~\citep{papineni2002bleu}. It measures the degree of n-gram overlapping between the string of each code snippet produced by the model and the reference, for values of $n$ usually ranging between $1$ and $4$ \citep{han2016machine,munkova2020evaluation}. This metric also takes into account a \textit{brevity penalty} to penalize predictions shorter than the references.
%BP is $1$ when the model output has (at least) the same length as the reference, and a decaying exponential when it is shorter. 
BLEU value ranges between $0$ and $1$, with higher scores corresponding to a better quality of the prediction. 

\vspace{0.05cm}
\noindent
$\blacksquare$ \textbf{Exact Match accuracy (EM)}. It indicates whether each code snippet produced by the model perfectly matches the reference. EM value is $1$ when there is an exact match, $0$ otherwise.

\vspace{0.05cm}
\noindent
$\blacksquare$ \textbf{METEOR}\citep{10.5555/1626355.1626389}. It measures the \textit{alignment} between each code snippet produced by the model and the reference. The alignment is defined as a mapping between unigrams (i.e., $1$-gram), such that every unigram in each string maps to zero or one unigram in the other string, and no unigrams in the same string. METEOR value ranges between $0$ and $1$, with higher scores corresponding to greater alignment between strings.

\vspace{0.05cm}
\noindent
$\blacksquare$ \textbf{Edit Distance (ED)}. It measures the \textit{edit distance} between two strings, i.e., the minimum number of operations on single characters required to make each code snippet produced by the model equal to the reference. ED value ranges between $0$ and $1$, with higher scores corresponding to smaller distances. 
\vspace{0.1cm}

%\noindent
%\textbf{Longest Common Subsequence (LCS)}. It measures the normalized similarity by calculating the longest common sub-sequence between the string of each code snippet produced by the model and the reference. 
%\vspace{0.1cm}

%\noindent
%\textbf{CodeBLEU}\citep{DBLP:journals/corr/abs-2009-10297}. It is defined as the weighted combination of four components: standard BLEU score, the weighted n-gram match (obtained by comparing the hypothesis code and the reference code tokens with different weights), the syntactic AST match (exploring the syntactic information of code), and the semantic dataflow match, considering the semantic similarity between the hypothesis and the reference. CodeBLEU value ranges between $0$ and $1$, with higher scores corresponding to better results.
%\vspace{0.1cm}

%\revisionA{\textbf{R1.1:} prova }

\subsection{Motivating Examples}
\label{subsec:motivating}

Output similarity metrics cannot properly assess whether two pieces of code are different but semantically equivalent, i.e., they provide the same output and/or effects although they use different operations (e.g., \texttt{jz label} and \texttt{je label} are different assembly instructions performing the same conditional jump).
For this reason, \textit{human evaluation} is considered the golden standard for assessing the quality of the code generated by the models. Through manual inspection of the model's predictions, human evaluation allows assessing the deeper linguistic features of the code~\citep{han-etal-2021-translation}, such as the  \textit{code semantics}, i.e., \textit{what the code actually does}. 
Therefore, for every code snippet generated by the model, we manually assess the \textbf{Semantic Correctness (SC)} metric, which indicates whether the output is the exact translation of the NL intent into the target programming language. This evaluation does not take into account the ground truth reference, but only the code predicted by the model and NL intent.
\textit{SC} value is either $1$ when the generated snippet is the (semantically) correct translation of the intent, and $0$ otherwise.
%Semantic correctness implies syntax correctness, whereas a snippet can be syntactically correct but semantically incorrect. Of course, the syntactic incorrectness also implies the semantic one~\citep{liguori2021evil}.
%As a simple example, consider the intent ‘‘increment the eax register'' and its reference assembly translation \texttt{inc eax}. If the model generates the snippet \texttt{add eax, 1}, then the output can be considered syntactically and semantically correct. Differently, if the output predicted by the model is, for example, \texttt{inc ebx}, then the snippet is still syntactically correct, but it fails to semantically match the reference. Note that this type of evaluation is rigorous. 
%\todo[noline,size=\footnotesize]{\textbf{Rev 2.1}}\revision{Unlike regular code generation tasks that focus on logically complex functional code fragments, code written for offensive purposes aims to gain full control over the memory layout and CPU registers of the target victim machine. This is usually accomplished through a large number of low-level arithmetic, logic operations, and bit-level slices.}

%\todo[noline,size=\footnotesize]{\textbf{Rev 2.1}}\revision{
Unlike regular code generation tasks that focus on logically complex functional code fragments, high-level offensive code often contains a large number of low-level arithmetic, logic operations, and bit-level slices to hide plain text attacks from antivirus and intrusion detection systems.%}

As a simple example, consider the intent ``\textit{compare string s1 with string s2}", which translates to the Python snippet:
\begin{center}
    \texttt{if s1 == s2:}
\end{center}
A semantically equivalent implementation of this string comparison is the code:
\begin{center}
\texttt{if s1.\_\_eq\_\_(s2):}
\end{center}
Despite the model's prediction being semantically correct ($SC = 1$), output similarity metrics are not able to grasp the equivalence between the two snippets since they base their calculation on character and/or token similarity. Therefore, this translation results in low automatic scores, including \textit{ROUGE-L} (P: $0.66$ R: $0.5$, F$_1$: $0.57$) and \textit{edit distance} ($0.47$).

The opposite occurs with the intent ``\textit{check if count modulo 2 is different from zero}'', which translates to the Python snippet:
\begin{center}
\texttt{if count \% 2 != 0:}
\end{center}
If the model generates the snippet: 
\begin{center}
\texttt{if count \% 2 == 0:}
\end{center}
then prediction and reference differ by a single character, yet the code accomplishes the opposite task. Automatic metrics fail to account for situations like this. For instance, the \textit{edit distance} between these two pieces of code is $0.94$, while the \textit{ROUGE-L} (P, R, F$_1$) score is $0.83$, which are considered high values. Differently, a human evaluator would appropriately classify this snippet as semantically incorrect (i.e., $SC = 0$), since it does not perform the intended check.  

%\todo[noline,size=\footnotesize]{\textbf{Rev 2.1}}\revision{
Furthermore, attackers leverage low-level programming languages, such as assembly, to perform surgically crafted exploitation of the system's low-level mechanisms, including heap metadata and stack return addresses, that are not accessible through high-level programming languages.
An example of an operation involving CPU registers is the intent ``\textit{transfer EAX contents into EDX register}", which translates to the assembly snippet:%}
\begin{center}
    \texttt{mov EDX, EAX}
\end{center}
An alternative method to copy the contents of a register into another is by pushing and popping its value onto the stack. Therefore, a semantically equivalent implementation of this copy is the code:
\begin{center}
\begin{tabular}{l}
     \texttt{push EAX} \\ 
     \texttt{pop EDX}
\end{tabular}
\end{center}

This translation is semantically correct when assessed by a human evaluator ($SC = 1$), yet it results in low scores for output similarity metrics such as \textit{ROUGE-L} ( F$_1$: $0.25$), \textit{BLEU-4} ($0.11$), \textit{edit distance} ($0.31$) and \textit{METEOR} ($0.24$).

Contrarily, there are situations in which the difference of a single character implies the use of a different register and, therefore, the implementation of a similar yet not equivalent operation. Indeed, consider the code description ``\textit{clear the EDX register and move 5 in the lowest byte of the register}'', which can be implemented through the assembly snippet:
\begin{center}
\begin{tabular}{l}
    \texttt{xor EDX, EDX} \\ 
    \texttt{mov DL, 5}
\end{tabular}
\end{center}
If the model generates the following code: 
\begin{center}
\begin{tabular}{l}
    \texttt{xor EDX, EDX} \\
    \texttt{mov BL, 5}
\end{tabular}
\end{center}

then the semantic correctness score should be zero ($SC = 0$) since the lowest byte of EDX is stored in the DL register, while BL contains the lowest byte of EBX. However, these two snippets are textually similar, hence resulting in high scores for \textit{edit distance} ($0.96$) and \textit{ROUGE-L} F$_1$ ($0.86$).

%% file: case_study.tex
\subsection{NMT Models}
To perform the code generation task, we consider two standard architectures: Seq2Seq, and CodeBERT.

\noindent
$\blacksquare$ \textbf{Seq2Seq} is a model that maps an input of sequence to an output of sequence. 
Similar to the encoder-decoder architecture with attention mechanism \citep{bahdanau2014neural}, we use a bi-directional LSTM as the encoder to transform an embedded intent sequence into a vector of hidden states with equal length. 
We implement the Seq2Seq model using \textit{xnmt}~\citep{neubig-etal-2018-xnmt}. 
We use an Adam optimizer \citep{kingma2015adam} with $\beta_1=0.9$ and $\beta_2=0.999$, while the learning rate $\alpha$ is set to $0.001$. We set all the remaining hyper-parameters in a basic configuration: layer dimension = $512$, layers = $1$, epochs = $200$, beam size = $5$.

\noindent
$\blacksquare$ \textbf{CodeBERT}~\citep{feng2020codebert} is a large multi-layer bidirectional Transformer architecture~\citep{vaswani2017attention} pre-trained on millions of lines of code across six different programming languages. 
%Like Seq2Seq, the Transformer architecture is made up of encoders and decoders. %CodeBERT has 12 stacked encoders and 6 stacked decoders. 
Our implementation uses an encoder-decoder framework where the encoder is initialized to the pre-trained CodeBERT weights, and the decoder is a transformer decoder, composed of $ 6$ stacked layers. The encoder follows the RoBERTa architecture~\citep{DBLP:journals/corr/abs-1907-11692}, with $12$ attention heads,  hidden layer dimension of $768$, $12$ encoder layers, and $514$ for the size of position embeddings. We set the learning rate $\alpha = 0.00005$, batch size = $32$, and beam size = $10$.

During data pre-processing, we tokenize the NL intents using the \textit{nltk word tokenizer}~\citep{bird2006nltk} and code snippets using the Python \textit{tokenize} package~\citep{tokenize}. 
We use \emph{spaCy}, an open-source, NL processing library written in Python and Cython~\citep{spacy}, to implement the named entity tagger for the standardization of the NL intents.

\subsection{Datasets}

\begin{table}[t]
\caption{Datasets statistics}
\label{tab:dataset_statistics}
\centering
\footnotesize
\begin{tabular}{
>{\centering\arraybackslash}m{5cm} |
>{\centering\arraybackslash}m{3cm}
>{\centering\arraybackslash}m{3cm}}
\toprule
\textbf{Statistic} & \textbf{Assembly Dataset} & \textbf{Python Dataset}\\ \midrule
\textit{Dataset size}  & $3,715$ & $15,540$\\ 
\textit{Unique Snippets}  &  $2,542$ & $14,034$\\ 
\textit{Unique Intents}  & $3,689$ & $15,421$\\
\textit{Unique tokens (Snippets)}  &  $1,657$ & $9,511$\\ 
\textit{Unique tokens (Intents)}  & $1,924$ & $10,605$\\ 
\textit{Avg. tokens per Snippet} & $4.75$ & $11.90$ \\
\textit{Avg. tokens per Intent}  & $9.53$ & $14.90$\\ \bottomrule
\end{tabular}
\end{table}

We feed the NMT models with a large corpus developed by~\cite{liguori2021evil} and used for security code generation and summarization~\citep{yang2022dualsc,yang2023exploitgen}.
The dataset fits in the context of the software security as it contains code snippets (alongside with their descriptions in English) of code used to develop and execute \textit{shellcode} programs, i.e., piece of code used as the payload in the exploitation of a software vulnerability. Examples of complex shellcode attacks include fork bombs, denial of service, bind shells, etc.

%This dataset represents an interesting case study because of its relevance in the software security field (code injection attacks are among the most common attacks nowadays~\citep{owasp}) and because
The dataset enables different analyses as it contains different programming languages.
Indeed, the corpus consists of two parts: (i) a \textit{Python dataset}, which contains Python code used by exploits to encode shellcodes (i.e., to obfuscate the execution of the shellcodes from anti-virus and intrusion detection systems), and (ii) an \textit{assembly dataset}, which includes shellcodes and decoders to revert the encoding. 
The authors collected exploits from publicly available databases, public repositories (e.g., GitHub), and programming guidelines.
A sample in the dataset consists of a snippet of code from these exploits and their corresponding description in the English language, as shown in \tablename{}~\ref{tab:dataset_instructions}.
The assembly dataset includes $783$ lines (${\sim}21\%$ of the dataset) of \emph{multi-line intents}, i.e., intents that generate multiple lines of assembly code (between $2$ and $5$), separated by the newline character \textit{\textbackslash{n}} (as the example in \tablename{}~\ref{tab:dataset_instructions}). %These multi-line snippets contain a number of assembly instructions that can range between $2$ and $5$. 

Table~\ref{tab:dataset_statistics} summarizes the statistics of both datasets, including the size (i.e., the unique pairs of intents-snippets), the unique lines of code snippets, the unique lines of NL intents, the unique number of tokens (i.e., words), and the average number of tokens per snippet and intent. The statistics highlight the difference between the datasets.
Unsurprisingly, the difference in terms of tokens per snippet ($\sim 12$ for Python, $\sim 5$ for assembly) and per intent ($\sim 15$ for Python, $\sim 10$ for assembly) makes the code generation task more difficult and challenging for the Python dataset.
Indeed, a low-level programming language contains a more limited amount of instructions and operations than a high-level language such as Python.
This explains why the size of the Python data is larger than the assembly data, i.e., to train the models to generate more complex programming code.
These differences allow us to evaluate the performance of the NMT models in offensive code generation with different complexity.

%Thus, we use distinct test sets for evaluating Python encoders and assembly language decoders, respectively.
%The test set for the Python data contains $375$ unique pairs of NL intents/Python snippets, while the test set of the assembly data contains $305$ unique pairs of NL intents/assembly snippets.

%Pearson correlation coefficient (PCC) is a statistical metric that measures the strength and direction of a linear relationship between two random variables~\citep{pearson1896vii}. It has been widely used in many applications, including to assess the correlation between the automatic metrics and the human evaluation in the code generation task~\citep{hu2022correlating}, pattern recognition~\citep{duda1973pattern}, time-delay estimation~\citep{xu2007time}, etc.

\begin{comment}
\subsection{Testbed}
We performed our experiments on a Linux OS running on a virtual machine. Seq2Seq utilized $8$ CPU cores and $8$ GB RAM. CodeBERT utilized 8 CPU cores, 16 GB RAM, and 2 GTX1080Ti GPUs. The computational time needed to generate the output depends on the settings of the hyper-parameters and the size of the dataset. 
On average, the training time of the Seq2Seq model is ${\sim}500$ minutes for the Python dataset and ${\sim}60$ minutes for the assembly dataset, while CodeBERT requires for the training in average ${\sim}220$ minutes for Python and ${\sim}25$ minutes for the assembly dataset.
\end{comment}

\begin{table}[t]
\caption{Examples of samples of the assembly and Python datasets}
\label{tab:dataset_instructions}
\centering
\footnotesize
\begin{tabular}
{ >{\centering\arraybackslash}m{2cm}|
>{\centering\arraybackslash}m{5cm} | 
>{\centering\arraybackslash}m{5cm}}
\toprule
\textbf{Dataset} & \textbf{NL Intent} & \textbf{Code Snippet} \\ \midrule
\textit{Assembly} & \textit{Copy the ASCII string \textit{``/bin//sh"} into the EBX register} & \texttt{push 0x68732f2f} \textbackslash{n} \texttt{push 0x6e69622f} \textbackslash{n} \texttt{mov ebx, esp} \\ \midrule
\textit{Python} & \textit{val2 is the result of the bitwise xor between the integer base 16 of the element i of chunk encoded to hex and xor\_byte} & \texttt{val2 = int( chunk[i]. encode('hex'), 16 ) \^{} xor\_byte}\\ \bottomrule
\end{tabular}
\end{table}

\subsection{Implementation of the Metrics}
To automatically assess the quality of the generated code snippets in comparison with the ground truth, we relied on open-source tools and Python packages. To compute the \textit{compilation accuracy}, we used the \textit{Netwide Assembler} (NASM) assembler~\citep{nasm} for the assembly code and the \texttt{py\_compile}~\citep{pycompile} compiler for Python snippets. As for the \textit{ROUGE} and \textit{ROUGE-L} metrics, we computed the output similarity scores using the Python package \texttt{rouge}~\citep{rouge} for both languages. We implemented \textit{BLEU} score computation employing the \texttt{bleu\_score} module contained in the open-source Python suite \textit{Natural Language Toolkit} (NLTK)~\citep{bleu}. For the \textit{edit distance} we used \texttt{pylcs}~\citep{pylcs}. To calculate the \textit{METEOR} metric, we relied on the Python library \texttt{evaluate} by HuggingFace~\citep{meteor}. Finally, for \textit{exact match accuracy}, we used a simple Python string comparison.

\subsection{Human Evaluation}
%\todo[noline,size=\footnotesize]{\textbf{Rev 3.2}}\revision{
To assess the semantic correctness of the predictions, we manually analyze every code snippet generated by the models to inspect if it is the correct translation of the NL intent, i.e., if the code snippet generated by the model performs what the English comment states. This analysis cannot be performed automatically (e.g., by comparing the predictions with ground-truth references) since an English intent can be translated into different but equivalent code snippets (as shown in \S{}~\ref{subsec:motivating}). 
This analysis can be subjective, as different reviewers may have different interpretations of the code and its intended functionality, depending on the expertise and experience of the reviewer. This can lead to inconsistent assessments of code correctness.
Moreover, manual analysis is prone to human error, as reviewers may miss subtle errors or inconsistencies in the code, or may introduce errors and biases into their assessments due to factors such as fatigue, distractions, or personal opinions.
Therefore, to reduce the possibility of errors and inconsistency, multiple authors performed the manual analysis independently and discussed the cases of discrepancy, obtaining a consensus for the semantic correctness of the predictions.%}

%% file: experiments.tex
Experiments aim to assess what are the output similarity metrics that are closer to the human evaluation in the generation of assembly and Python code. 
To perform the experiments, we split the dataset into training (the set of examples used to fit the parameters), validation (the set used to tune the hyperparameters of the models), and test (the set used for the evaluation of the models) sets using a common $80\%/10\%/10\%$ ratio~\citep{kim2018artificial,DBLP:conf/msr/MashhadiH21,liguori-etal-2021-shellcode}. 
%The test sets used for the evaluation cover $20$ different exploits (i.e., $20$ assembly programs and $20$ Python programs), including $305$ and $375$ unique pairs of NL intents/code snippets of assembly and Python code, respectively.

\subsection{Quantitative Analysis}

\begin{figure}
    \centering
    \includegraphics[width=0.75\columnwidth]{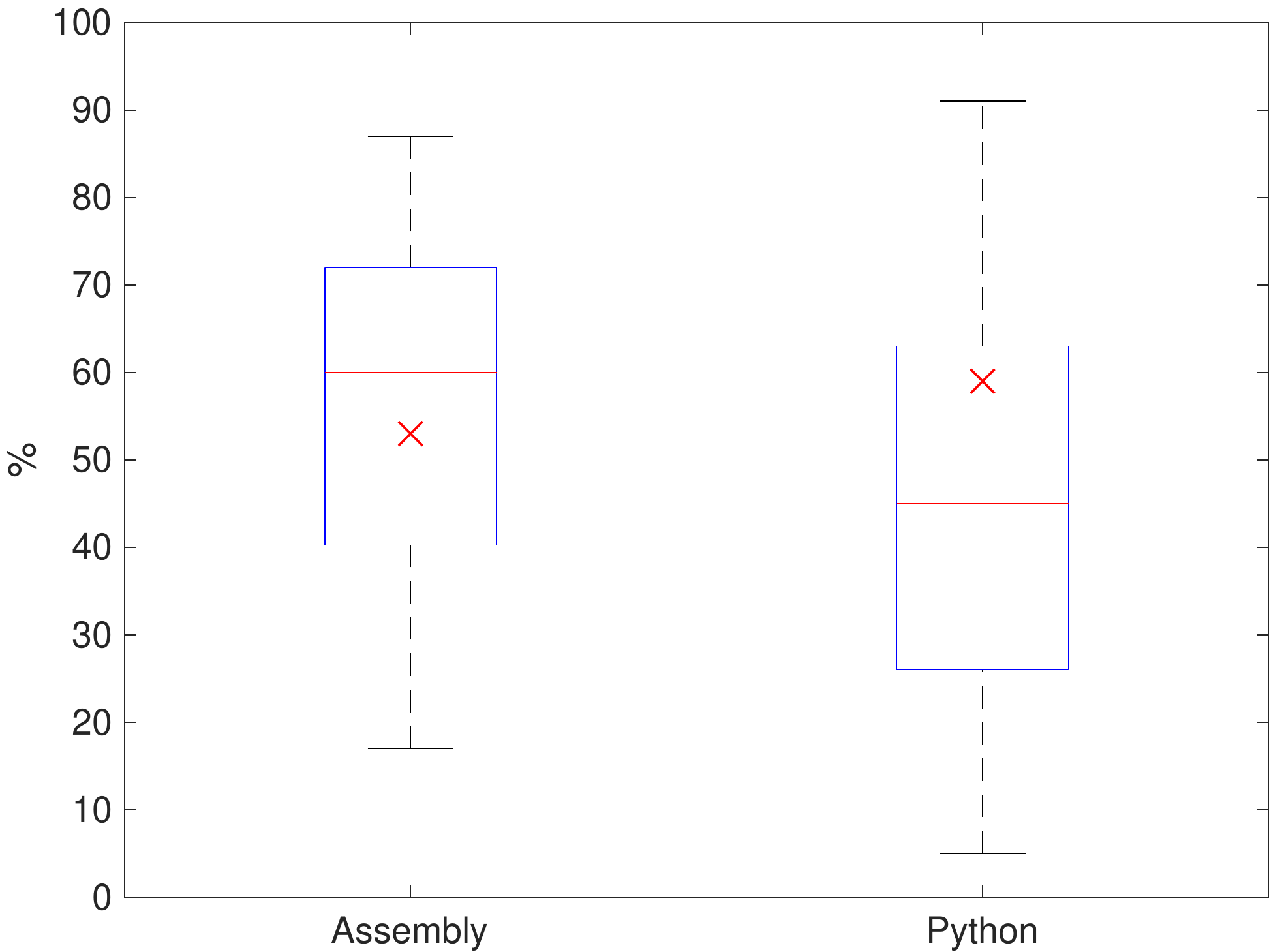}
    \caption{Distribution of the output similarity metrics on the assembly and Python data. The marker \textbf{\textcolor{red}{X}} indicates the semantic correctness value.}
    \label{fig:boxplot}
\end{figure}

First, we investigated whether the output similarity metrics provide similar results on the same data. To this aim, we assessed the performance of the NMT models on the assembly and Python test data. \figurename{}~\ref{fig:boxplot} shows the statistics, in terms of boxplots, of the output similarity metrics. The height of both boxplots highlights that the metrics provide very different results. 
Indeed, we found that for the assembly data, the min value is $17\%$, the median is $60\%$ (the average is $56\%$, the standard deviation is $21\%$), and the max value is $87\%$. 
For the Python data, the min value is $5\%$, the median is $45\%$ (the average is $46\%$, the standard deviation is $24\%$) and the max is $91\%$. Therefore, different metrics used for code generation provide very different values, leading to a wrong assessment of the performance of the models.

The figure also shows the SC values (\textbf{\textcolor{red}{X}} markers), which are $53\%$ and $59\%$ for assembly and Python data, respectively (i.e., the data is pretty balanced). Ideally, output similarity metrics should provide a value closer to the one of semantic correctness.
We found that the median value of the output similarity metrics is $7\%$ higher than the human evaluation for assembly data, while is $14\%$ lower than the human evaluation for Python data, i.e., the metrics overestimate the performance for the assembly and underestimate the performance for the Python data.
We attribute this behavior to the different structures of the assembly and Python code. Indeed, the limited set of instructions of a low-level language makes the code snippets similar even when they are semantically equivalent (e.g., \texttt{jz label} and \texttt{jnz label} are similar but semantically different instructions). This is not the case for the Python dataset, which has a higher complexity in terms of different instructions, length of the code, etc. Therefore, it is more likely to write equivalent code with very different instructions.

%First, we compared the value of the output similarity metrics with the human evaluation on the assembly and Python test sets. 
We then performed an in-depth analysis by comparing each output similarity metric with the human evaluation, i.e., the SC.
Specifically, for each dataset, we performed three different analyses, depending on the code snippets included in the test set for the evaluation: i) the \textit{whole test set}, i.e, we considered all the code snippets in the test data, ii) only the \textit{correct predictions}, i.e., we limited the analysis to the code snippets considered correct according to the human evaluation, and iii) only the \textit{wrong predictions}, i.e., the analysis entailed only the code snippets considered as semantically incorrect by the human evaluation. The analysis of the correct predictions and wrong predictions enables different considerations, e.g., to infer what are the most suitable metrics to use when the models provide very accurate predictions or when they deal with very challenging code generation tasks.
To this aim, we computed an \textit{offset value}, i.e., the distance between the optimal value represented by the human evaluation (i.e., the semantic correctness) and the estimate provided by the output similarity metrics: the lower the offset, the closer the automatic metric is to the human evaluation.

\begin{table}[ht!]
\caption{Average values and offset of the output similarity metrics on the \textbf{assembly} dataset. Best performance (lower offset) is \textcolor{blue}{blue}, while the worst performance (higher offset) is \textcolor{red}{red}.}
\label{tab:assembly_corpus}
\centering
\footnotesize
\begin{tabular}{
>{\arraybackslash}m{2.5cm} |
>{\arraybackslash}m{1.25cm}  
>{\arraybackslash}m{1.25cm} |
>{\arraybackslash}m{1.25cm} 
>{\arraybackslash}m{1.25cm} |
>{\arraybackslash}m{1.25cm}
>{\arraybackslash}m{1.25cm}} 
\toprule
\centering\textit{\textbf{Metric}}           & \multicolumn{2}{c}{\textit{\textbf{Whole}}} & \multicolumn{2}{c}{\textit{\textbf{Correct}}} & \multicolumn{2}{c}{\textit{\textbf{Wrong}}} \\ 
 & Value & Offset  & Value & Offset  & Value & Offset \\ \midrule
\textit{SC} & 0.53 & - & 1.00 & - & 0.00 & - \\ \midrule 
\textit{CA} & 0.87& \Chart{0.34} & \textcolor{blue}{1.00} & \textcolor{blue}{\Chart{0.00}} & \textcolor{red}{0.72} & \textcolor{red}{\Chart{0.72}} \\
\textit{ROUGE-1 P} & 0.75& \Chart{0.22} & 0.93& \Chart{0.07} & 0.55& \Chart{0.55} \\
\textit{ROUGE-1 R} & 0.71& \Chart{0.18} & 0.92& \Chart{0.08} & 0.47& \Chart{0.47} \\
\textit{ROUGE-1 $F_1$} & 0.72& \Chart{0.19} & 0.92& \Chart{0.08} & 0.50& \Chart{0.50} \\
\textit{ROUGE-2 P} & 0.55& \Chart{0.02} & 0.75& \Chart{0.25} & 0.32& \Chart{0.32} \\
\textit{ROUGE-2 R} & 0.52& \Chart{0.01} & 0.74& \Chart{0.26} & 0.28& \Chart{0.28} \\
\textit{ROUGE-2 $F_1$} & \textcolor{blue}{0.53} & \textcolor{blue}{\Chart{0.00}} & 0.74 & \Chart{0.26} & 0.28& \Chart{0.28} \\
\textit{ROUGE-3 P} & 0.40& \Chart{0.12} & 0.60& \Chart{0.40} & 0.19& \Chart{0.19} \\
\textit{ROUGE-3 R} & 0.38& \Chart{0.14} & 0.59& \Chart{0.41} & 0.16& \Chart{0.16} \\
\textit{ROUGE-3 $F_1$} & 0.39& \Chart{0.14} & 0.59& \Chart{0.41} & 0.16& \Chart{0.16} \\
\textit{ROUGE-4 P} & 0.18& \Chart{0.35} & \textcolor{red}{0.25} & \textcolor{red}{\Chart{0.75}} & 0.10 & \Chart{0.10} \\
\textit{ROUGE-4 R} & \textcolor{red}{0.17} & \textcolor{red}{\Chart{0.36}} & \textcolor{red}{0.25} & \textcolor{red}{\Chart{0.75}} & 0.09 & \Chart{0.09} \\
\textit{ROUGE-4 $F_1$} & \textcolor{red}{0.17} & \textcolor{red}{\Chart{0.36}} & \textcolor{red}{0.25} & \textcolor{red}{\Chart{0.75}} & 0.09 & \Chart{0.09} \\
\textit{ROUGE-L P} & 0.75& \Chart{0.22} & 0.93& \Chart{0.07} & 0.54& \Chart{0.54} \\
\textit{ROUGE-L R} & 0.71& \Chart{0.18} & 0.92& \Chart{0.08} & 0.47& \Chart{0.47} \\
\textit{ROUGE-L $F_1$} & 0.72& \Chart{0.19} & 0.92& \Chart{0.08} & 0.49& \Chart{0.49} \\
\textit{BLEU-1} & 0.69& \Chart{0.16} & 0.89& \Chart{0.11} & 0.53& \Chart{0.53} \\
\textit{BLEU-2} & 0.63& \Chart{0.10} & 0.86& \Chart{0.14} & 0.45& \Chart{0.45} \\
\textit{BLEU-3} & 0.60& \Chart{0.07} & 0.84& \Chart{0.16} & 0.41& \Chart{0.41} \\
\textit{BLEU-4} & 0.57 & \Chart{0.04} & 0.81& \Chart{0.19} & 0.39& \Chart{0.39} \\
\textit{EM} & 0.41& \Chart{0.11} & 0.79& \Chart{0.21} & \textcolor{blue}{0.00} & \textcolor{blue}{\Chart{0.00}} \\
\textit{METEOR} & 0.72& \Chart{0.19} & 0.90& \Chart{0.10} & 0.52& \Chart{0.52} \\
\textit{ED} & 0.79& \Chart{0.27} & 0.96& \Chart{0.04} & 0.60& \Chart{0.60} \\\midrule
\textit{\textbf{Average}} & 0.56	& \Chart{0.17}	& 0.75 & \Chart{0.25} & 0.36 &	\Chart{0.36}\\
\bottomrule
\end{tabular}
\end{table}

\begin{table}[ht!]
\centering
\caption{Average values and offset of the output similarity metrics on the \textbf{Python} dataset. Best performance (lower offset) is \textcolor{blue}{blue}, while the worst performance (higher offset) is \textcolor{red}{red}.}
\label{tab:python_corpus}
\centering
\footnotesize
\begin{tabular}{
>{\arraybackslash}m{2.5cm} |
>{\arraybackslash}m{1.25cm}  
>{\arraybackslash}m{1.25cm} |
>{\arraybackslash}m{1.25cm} 
>{\arraybackslash}m{1.25cm} |
>{\arraybackslash}m{1.25cm}
>{\arraybackslash}m{1.25cm}} 
\toprule
\centering\textit{\textbf{Metric}}           & \multicolumn{2}{c}{\textit{\textbf{Whole}}} & \multicolumn{2}{c}{\textit{\textbf{Correct}}} & \multicolumn{2}{c}{\textit{\textbf{Wrong}}} \\ 
 & Value & Offset  & Value & Offset  & Value & Offset \\ \midrule
\textit{SC} & 0.59 & - & 1.00 & - & 0.00 & - \\ \midrule 
\textit{CA} & 0.91  & \Chart{0.32} & \textcolor{blue}{1.00} & \textcolor{blue}{\Chart{0.00}} & \textcolor{red}{0.79}  & \textcolor{red}{\Chart{0.79}}\\
\textit{ROUGE-1 P} & 0.63  & \Chart{0.04} & 0.75  & \Chart{0.25} & 0.46  & \Chart{0.46}\\
\textit{ROUGE-1 R} & 0.63  & \Chart{0.04} & 0.74  & \Chart{0.26} & 0.47  & \Chart{0.47}\\
\textit{ROUGE-1 $F_1$} & 0.63  & \Chart{0.04} & 0.74  & \Chart{0.26} & 0.46  & \Chart{0.46}\\
\textit{ROUGE-2 P} & 0.45  & \Chart{0.14} & 0.59  & \Chart{0.41} & 0.23  & \Chart{0.23}\\
\textit{ROUGE-2 R} & 0.45  & \Chart{0.14} & 0.58  & \Chart{0.42} & 0.24  & \Chart{0.24}\\
\textit{ROUGE-2 $F_1$} & 0.44  & \Chart{0.15} & 0.58  & \Chart{0.42} & 0.23  & \Chart{0.23}\\
\textit{ROUGE-3 P} & 0.26  & \Chart{0.33} & 0.38  & \Chart{0.62} & 0.07  & \Chart{0.07}\\
\textit{ROUGE-3 R} & 0.26  & \Chart{0.33} & 0.38  & \Chart{0.62} & 0.07  & \Chart{0.07}\\
\textit{ROUGE-3 $F_1$} & 0.26  & \Chart{0.33} & 0.38  & \Chart{0.62} & 0.07  & \Chart{0.07}\\
\textit{ROUGE-4 P} & \textcolor{red}{0.05}  & \textcolor{red}{\Chart{0.54}} & \textcolor{red}{0.07}  & \textcolor{red}{\Chart{0.93}} & 0.02  & \Chart{0.02}\\
\textit{ROUGE-4 R} & \textcolor{red}{0.05}  & \textcolor{red}{\Chart{0.54}} & \textcolor{red}{0.07}  & \textcolor{red}{\Chart{0.93}} & 0.02  & \Chart{0.02}\\
\textit{ROUGE-4 $F_1$} & \textcolor{red}{0.05}  & \textcolor{red}{\Chart{0.54}} & \textcolor{red}{0.07}  & \textcolor{red}{\Chart{0.93}} & 0.02  & \Chart{0.02}\\
\textit{ROUGE-L P} & 0.63  & \Chart{0.04} & 0.75  & \Chart{0.25} & 0.46  & \Chart{0.46}\\
\textit{ROUGE-L R} & 0.63  & \Chart{0.04} & 0.74  & \Chart{0.26} & 0.47  & \Chart{0.47}\\
\textit{ROUGE-L $F_1$} & 0.63  & \Chart{0.04} & 0.74  & \Chart{0.26} & 0.46  & \Chart{0.46}\\
\textit{BLEU-1} & \textcolor{blue}{0.58}  & \textcolor{blue}{\Chart{0.01}} & 0.69  & \Chart{0.31} & 0.44  & \Chart{0.44}\\
\textit{BLEU-2} & 0.48  & \Chart{0.11} & 0.61  & \Chart{0.39} & 0.32  & \Chart{0.32}\\
\textit{BLEU-3} & 0.37  & \Chart{0.22} & 0.50  & \Chart{0.50} & 0.19  & \Chart{0.19}\\
\textit{BLEU-4} & 0.26  & \Chart{0.33} & 0.37  & \Chart{0.63} & 0.13  & \Chart{0.13}\\
\textit{EM} & 0.27  & \Chart{0.32} & 0.43  & \Chart{0.57} & \textcolor{blue}{0.00}  & \textcolor{blue}{\Chart{0.00}}\\
\textit{METEOR} & 0.74  & \Chart{0.16} & 0.84  & \Chart{0.16} & 0.60  & \Chart{0.60}\\
\textit{ED} & 0.81  & \Chart{0.22} & 0.91  & \Chart{0.09} & 0.66  & \Chart{0.66}\\ \midrule
\textit{\textbf{Average}} & 0.46 &	\Chart{0.22} & 0.56 & \Chart{0.44} & 0.30 & \Chart{0.30}\\
\bottomrule
\end{tabular}
\end{table}

\tablename{}~\ref{tab:assembly_corpus} and \tablename{}~\ref{tab:python_corpus} show the results, including the average values obtained by the metrics and the offset with the \textit{SC}, for the assembly and Python datasets, respectively. %For the whole test set evaluation, we first notice that data is pretty balanced, as we have $53\%$ and $59\%$ of semantically correct predictions for assembly and Python datasets.
In this analysis, we found that the metric closer to the semantic correctness is \textit{ROUGE-2} (offset is $0\%$, i.e., they provide the same value) for the assembly language, while \textit{BLEU-1} is closer to the human evaluation on the Python code snippets (offset = $1\%$). 
For both assembly and Python code snippets, the metric less similar to the \textit{SC} is \textit{ROUGE-4} (offset = $36\%$ and $54\%$, respectively). Therefore, metrics based on n-grams provide an evaluation close to the semantic correctness when $n$ is equal to $1$ or $2$, while their ability in the evaluation gets worse when $n \geq 3$.
We attribute this behavior to the brevity (in terms of tokens) of programming language snippets, especially in the context of the generation of software exploits, which require low-level instructions and binary-level-data processing. In this case, since code frequently includes operations with few tokens (e.g., a \texttt{jmp label} instruction, a Python increment with $\mathrel{+}=$ operand), output similarity metrics using a higher number of n-grams underestimate the quality of the predictions (e.g., \textit{ROUGE-4} for assembly and Python, \textit{BLEU-4} for Python).  
%Other metrics, such as \textit{compilation accuracy}, \textit{METEOR}, \textit{exact match}, and \textit{edit distance}, are still too far from the \textit{SC} values.

When we limit the evaluation to the correct predictions, i.e., we evaluate only the code snippets considered semantically correct ($SC=1$), then the best metric is the \textit{compilation accuracy}, regardless of the programming language. Indeed, when the snippet is semantically correct then it is also syntactically correct, i.e., it is also compilable. Therefore, the offset is $0\%$ in this case. Besides the \textit{CA}, the \textit{edit distance} is a valuable option in this specific case (offsets are $4\%$ and $9\%$ for assembly and Python datasets) since the number of operations required to make the predictions equal to the reference is limited when the predictions are correct.
The worst metric, instead, is again \textit{ROUGE-4} (offset equal to $75\%$ and $93\%$ for assembly and Python, respectively), showing that it is not able to properly assess the correctly generated code snippets.

The analysis on the wrong predictions, i.e., on the code snippets not semantically correct ($SC=0$), highlights that the \textit{exact match} accuracy is a good evaluator for both datasets as the models' predictions, in this case, never match the ground truth references (and, therefore, the values are $0$ for every code snippet). 
The \textit{compilation accuracy}, which showed the best performance in the previous case study, provides the worst performance on the wrong predictions since a semantically incorrect snippet can be syntactically correct (i.e., compilable). Indeed, we found that $72\%$ for the assembly dataset and $79\%$ for the Python dataset of the semantically incorrect snippets are syntactically correct. 

Finally, we found that the average values of the offsets over all the code similarity metrics are pretty similar for both assembly and Python datasets for the whole test set ($17\%$ vs $22\%$) and the wrong predictions ($36\%$ vs $30\%$). For the correct predictions, instead, the differences between assembly and Python are more exacerbated ($25\%$ vs $44\%$) due to the ability of the high-level language to write semantically equivalent code with different instructions.

\subsection{Correlation Analysis}
\label{subsec:correlation_analysis}
We further assessed the ability of the output similarity metrics in the evaluation of the code generation task. Different from the previous quantitative analysis, in which we compared the average values provided by the output similarity metrics with the average semantic correctness over all the test sets, we performed an in-depth statistical analysis by computing the correlation of the output similarity metrics with the human evaluation of all the code snippets of the assembly and Python test sets (i.e., we considered the values of the metrics on the single predictions).

%We computed the \textit{Pearson} correlation coefficient and \textit{Kendall} rank correlation coefficient to correlate the automatic metrics with the semantic correctness. 
We computed the \textit{Pearson} correlation coefficient $r$, which measures the strength of association (i.e., the linear relationship) between two variables in a correlation analysis and is defined as the covariance of the two variables divided by the product of their respective standard deviations~\citep{pearson1895notes}.
Moreover, to assess the relationship between the metrics and the semantic correctness, we also computed the \textit{Kendall} correlation coefficient $\tau$, which measures the dependence between two random variables based on the ranks of sampled observations of the variable~\citep{kendall1938new}. %It makes use of the idea of concordance: two random variables are concordant if large (small) values of one are related to large (small) values of the other. When large (small) values of one are related to small (large) values of the other, the random variables are discordant.

The correlation coefficients are unit-free values between $-1$ and $1$, which represent \textit{perfect} correlations, \textit{negative}, and \textit{positive}, respectively.
Positive values indicate a positive correlation, i.e., the values of both variables tend to increase together, while negative values indicate a negative correlation, i.e., the values of one variable tend to increase when the values of the other variable decrease.
Therefore, a high value of the coefficient indicates that the output similarity metric is strongly associated with human evaluation. On the contrary, a small value indicates that the automatic metric is poorly associated with human evaluation.

\begin{table}[ht!]
\caption{Correlation Analysis between output similarity metrics and human evaluation on the \textbf{assembly} dataset. Best performance is \textcolor{blue}{blue}, while the worst performance is \textcolor{red}{red}.}
\label{tab:correlation_assembly}
\centering
\footnotesize
\begin{tabular}{
>{\arraybackslash}m{2.5cm} |
>{\arraybackslash}m{2cm}  
>{\arraybackslash}m{2cm} |
>{\arraybackslash}m{2cm}
>{\arraybackslash}m{2cm}}
\toprule
& \multicolumn{2}{c}{\textit{\textbf{Seq2Seq}}}         & \multicolumn{2}{c}{\textit{\textbf{CodeBERT}}} \\
\centering\textit{\textbf{Output Similarity Metric}}             & \textit{\textbf{Pearson's $r$}} & \textit{\textbf{Kendall's $\tau$}} & \textit{\textbf{Pearson's $r$}} & \textit{\textbf{Kendall's $\tau$}}\\ \midrule
\textit{CA} &	\Chart{0.35} &	\Chart{0.35} &	\Chart{0.47} &	\Chart{0.47}\\
\textit{ROUGE-1 P} &	\Chart{0.63} &	\Chart{0.63} &	\Chart{0.60} &	\Chart{0.62}\\
\textit{ROUGE-1 R} &	\Chart{0.70} &	\Chart{0.67} &	\Chart{0.65} &	\Chart{0.66}\\
\textit{ROUGE-1 $F_1$} &	\Chart{0.70} &	\Chart{0.67} &	\Chart{0.65} &	\Chart{0.66}\\
\textit{ROUGE-2 P} &	\Chart{0.49} &	\Chart{0.47} &	\Chart{0.53} &	\Chart{0.51}\\
\textit{ROUGE-2 R} &	\Chart{0.54} &	\Chart{0.49} &	\Chart{0.55} &	\Chart{0.52}\\
\textit{ROUGE-2 $F_1$} &	\Chart{0.54} &	\Chart{0.50} &	\Chart{0.55} &	\Chart{0.52}\\
\textit{ROUGE-3 P} &	\Chart{0.46} &	\Chart{0.42} &	\Chart{0.47} &	\Chart{0.44}\\
\textit{ROUGE-3 R} &	\Chart{0.51} &	\Chart{0.44} &	\Chart{0.47} &	\Chart{0.44}\\
\textit{ROUGE-3 $F_1$} &	\Chart{0.50} &	\Chart{0.44} &	\Chart{0.47} &	\Chart{0.44}\\
\textit{ROUGE-4 P} &	\textcolor{red}{\Chart{0.19}} &	\textcolor{red}{\Chart{0.12}} &	\textcolor{red}{\Chart{0.22}} &	\textcolor{red}{\Chart{0.17}}\\
\textit{ROUGE-4 R} &	\Chart{0.23} &	\Chart{0.13} &	\Chart{0.23} &	\Chart{0.18}\\
\textit{ROUGE-4 $F_1$} &	\Chart{0.22} &	\Chart{0.13} &	\Chart{0.23} &	\Chart{0.18}\\
\textit{ROUGE-L P} &	\Chart{0.63} &	\Chart{0.62} &	\Chart{0.62} &	\Chart{0.64}\\
\textit{ROUGE-L R} &	\Chart{0.70} &	\Chart{0.67} &	\Chart{0.67} &	\Chart{0.67}\\
\textit{ROUGE-L $F_1$} &	\Chart{0.69} &	\Chart{0.67} &	\Chart{0.66} &	\Chart{0.67}\\
\textit{BLEU-1} &	\Chart{0.72} &	\Chart{0.67} &	\Chart{0.66} &	\Chart{0.65}\\
\textit{BLEU-2} &	\Chart{0.56} &	\Chart{0.51} &	\Chart{0.56} &	\Chart{0.53}\\
\textit{BLEU-3} &	\Chart{0.50} &	\Chart{0.49} &	\Chart{0.48} &	\Chart{0.48}\\
\textit{BLEU-4} &	\Chart{0.22} &	\Chart{0.36} &	\Chart{0.23} &	\Chart{0.32}\\
\textit{EM} &	\textcolor{blue}{\Chart{0.81}} &	\textcolor{blue}{\Chart{0.81}} &	\textcolor{blue}{\Chart{0.78}} &	\textcolor{blue}{\Chart{0.78}}\\
\textit{METEOR} &	\Chart{0.68} &	\Chart{0.61} &	\Chart{0.61} &	\Chart{0.57}\\
\textit{ED} &	\Chart{0.72} &	\Chart{0.71} &	\Chart{0.68} &	\Chart{0.70}\\ \midrule
\textit{\textbf{Average}} & \Chart{0.53} & \Chart{0.50} & \Chart{0.52} &	\Chart{0.51}\\
\bottomrule
\end{tabular}
\end{table}

\begin{table}[ht!]
\caption{Correlation Analysis between output similarity metrics and human evaluation on the \textbf{Python} dataset. Best performance is \textcolor{blue}{blue}, while the worst performance is \textcolor{red}{red}.}
\label{tab:correlation_python}
\centering
\footnotesize
\begin{tabular}{
>{\arraybackslash}m{2.5cm} |
>{\arraybackslash}m{2cm}  
>{\arraybackslash}m{2cm} |
>{\arraybackslash}m{2cm}
>{\arraybackslash}m{2cm}}
\toprule
& \multicolumn{2}{c}{\textit{\textbf{Seq2Seq}}}         & \multicolumn{2}{c}{\textit{\textbf{CodeBERT}}} \\
\centering\textit{\textbf{Output Similarity Metric}}             & \textit{\textbf{Pearson's $r$}} & \textit{\textbf{Kendall's $\tau$}} & \textit{\textbf{Pearson's $r$}} & \textit{\textbf{Kendall's $\tau$}} \\ \midrule
\textit{CA} &	\Chart{0.36} &	\Chart{0.36} &	\Chart{0.38} &	\Chart{0.38} \\
\textit{ROUGE-1 P} &	\Chart{0.49} &	\Chart{0.41} &	\Chart{0.50} &	\Chart{0.46} \\
\textit{ROUGE-1 R} &	\Chart{0.47} &	\Chart{0.38} &	\Chart{0.45} &	\Chart{0.41} \\
\textit{ROUGE-1 $F_1$} &	\Chart{0.49} &	\Chart{0.40} &	\Chart{0.48} &	\Chart{0.43} \\
\textit{ROUGE-2 P} &	\Chart{0.46} &	\Chart{0.40} &	\Chart{0.48} &	\Chart{0.42} \\
\textit{ROUGE-2 R} &	\Chart{0.44} &	\Chart{0.37} &	\Chart{0.44} &	\Chart{0.39} \\
\textit{ROUGE-2 $F_1$} &	\Chart{0.46} &	\Chart{0.38} &	\Chart{0.46} &	\Chart{0.40} \\
\textit{ROUGE-3 P} &	\Chart{0.31} &	\Chart{0.23} &	\Chart{0.43} &	\Chart{0.40} \\
\textit{ROUGE-3 R} &	\Chart{0.30} &	\Chart{0.23} &	\Chart{0.42} &	\Chart{0.39} \\
\textit{ROUGE-3 $F_1$} &	\Chart{0.31} &	\Chart{0.23} &	\Chart{0.43} &	\Chart{0.40} \\
\textit{ROUGE-4 P} &	\Chart{0.15} &		\textcolor{red}{\Chart{0.12}} &	\Chart{0.10} &		\textcolor{red}{\Chart{0.04}} \\
\textit{ROUGE-4 R} &	\Chart{0.14} &		\textcolor{red}{\Chart{0.12}} &		\textcolor{red}{\Chart{0.09}} &		\textcolor{red}{\Chart{0.04}} \\
\textit{ROUGE-4 $F_1$} &	\Chart{0.15} &		\textcolor{red}{\Chart{0.12}} &	\Chart{0.10} &		\textcolor{red}{\Chart{0.04}} \\
\textit{ROUGE-L P} &	\Chart{0.50} &	\Chart{0.42} &	\Chart{0.51} &	\Chart{0.46} \\
\textit{ROUGE-L R} &	\Chart{0.48} &	\Chart{0.38} &	\Chart{0.45} &	\Chart{0.41} \\
\textit{ROUGE-L $F_1$} &	\Chart{0.50} &	\Chart{0.41} &	\Chart{0.49} &	\Chart{0.44} \\
\textit{BLEU-1} &	\Chart{0.43} &	\Chart{0.36} &	\Chart{0.48} &	\Chart{0.42} \\
\textit{BLEU-2} &	\Chart{0.45} &	\Chart{0.37} &	\Chart{0.47} &	\Chart{0.40} \\
\textit{BLEU-3} &	\Chart{0.33} &	\Chart{0.29} &	\Chart{0.44} &	\Chart{0.35} \\
\textit{BLEU-4} &	\textcolor{red}{\Chart{0.13}} &	\Chart{0.20} &		\textcolor{red}{\Chart{0.09}} &	\Chart{0.13} \\
\textit{EM} &	\Chart{0.42} &	\Chart{0.42} &	\Chart{0.54} &	\Chart{0.54} \\
\textit{METEOR} &	\Chart{0.38} &	\Chart{0.32} &	\Chart{0.55} &	\Chart{0.53} \\
\textit{ED} &	\textcolor{blue}{\Chart{0.57}} &	\textcolor{blue}{\Chart{0.49}} &	\textcolor{blue}{\Chart{0.57}} &	\textcolor{blue}{\Chart{0.56}} \\ \midrule
\textit{\textbf{Average}} & \Chart{0.38} & \Chart{0.32} &	\Chart{0.41} &	\Chart{0.37}\\
\bottomrule
\end{tabular}
\end{table}

\tablename{}~\ref{tab:correlation_assembly} and \tablename{}~\ref{tab:correlation_python} show Pearson's $r$ and Kendall's $\tau$ correlation coefficients between the automatic metrics and the semantic correctness on the assembly and Python datasets, respectively, for both the Seq2Seq and CodeBERT models.
For the assembly dataset, we found that the \textit{exact match} has the highest correlation coefficients for both models ($r$ and $\tau$ are $0.81$ for Seq2Seq and $0.78$ for CodeBERT). We attribute this result to the nature of the assembly language, which has a fixed structure and provides a more limited set of instructions to express an operation if compared to a high-level language. Therefore, the \textit{exact match}, which provides $1$ only when the prediction is equal to the reference, results to be the most suitable metric for this case study. 

For the Python dataset, the \textit{edit distance} is the most correlated metric with the human evaluation ($r = 0.57$ and $\tau = 0.49$ for Seq2Seq, while $r = 0.57$ and $\tau = 0.56$ for CodeBERT), although the coefficients are lower than the best values of the assembly case study.
%\textit{ROUGE-L} confirms to be a valuable alternative for the evaluation, regardless of the language of the code snippets. 
%Other metrics, such as METEOR and compilation accuracy, were revealed to be not very accurate in the evaluation of the semantic correctness of the code, regardless of the code generated by the models.
As above mentioned, Python code for software exploits requires a considerable amount of binary-level-data processing and concise instructions. If a single character in a hexadecimal value is not correct, then, for n-gram-based metrics such as ROUGE, the whole token is different from the reference, resulting in low scores. Differently, character-based metrics such as the \textit{edit distance} account for these slight deviations and result in a high score.

Again, \textit{ROUGE-4} metrics provide the worst results for both datasets. 
For the assembly language, the correlation coefficients are $r \leq 0.23$ and $\tau \leq 0.13$ for Seq2Seq, while $r \leq 0.23$ and $\tau \leq 0.18$ for CodeBERT. For the Python code snippets, the correlation coefficients are even lower ($r \leq 0.15$, $\tau \leq 0.12$). This result confirms that n-gram-based metrics are lowly correlated to the human evaluation when $n$ is high.

The comparison of the correlation coefficients on the assembly and Python datasets highlights that the code similarity metrics are more correlated to the assembly case study ($r \geq 0.52$, $\tau \geq 0.50$) than the Python one ($r \leq 0.41$, $\tau \leq 0.37$) due to the increasing difficulty of the code similarity metrics to assess the generation of more complex offensive code.

\begin{comment}
\begin{mybox}{Results of the correlation analysis.}
The results of the correlation analysis confirm that the exact match and the edit distance are the most suitable metrics to assess assembly and Python code snippets, respectively.
The n-gram-based metrics, such as ROUGE and BLEU scores, provide acceptable results when the number of n-grams is kept low (i.e., 1 or 2). The ROUGE-L confirms to be a valuable alternative for the evaluation, regardless of the language of the code snippets.
Other metrics, such as METEOR and compilation accuracy, were revealed to be not very accurate in the evaluation of the semantic correctness of the code, regardless of the code generated by the models.
\end{mybox}
\end{comment}

%% file: discussion.tex
\begin{table}[t]
\caption{Examples that show how different output similarity metrics work for different programming languages. \textcolor{red}{Red} refers to incorrect predictions.}
\label{tab:metrics_examples}
\centering
\footnotesize
\begin{tabular}
{ >{\centering\arraybackslash}m{1.4cm}|
>{\centering\arraybackslash}m{2.8cm} | 
>{\centering\arraybackslash}m{2.8cm} | 
>{\centering\arraybackslash}m{0.6cm} |
>{\centering\arraybackslash}m{1.85cm}| 
>{\centering\arraybackslash}m{0.6cm} |
>{\centering\arraybackslash}m{0.6cm}}
\toprule
\textbf{Dataset} & \textbf{Ground Truth} & \textbf{Predicted Code} & \textbf{\textit{SC}} & \textbf{\textit{ROUGE-4 (F$_1$)}} & \textbf{\textit{ED}} & \textbf{\textit{EM}}\\ \midrule

& \texttt{add EAX, EBX} & \texttt{add EAX, EBX} & $1.0$ & $0.0$ & $1.0$ & $1.0$\\ \cmidrule{2-7}

\textit{Assembly} & \texttt{xor ECX, ECX \textbackslash{n} mul ECX} & \texttt{xor ECX, ECX \textbackslash{n} mul \textcolor{red}{EBX}} & $0.0$ & $0.66$ & $0.95$ & $0.0$\\ \cmidrule{2-7}

& \texttt{jmp decode} & \texttt{jmp decode} & $1.0$ & $0.0$ & $1.0$ & $1.0$\\ \bottomrule

%& \texttt{x = x << 1} & \texttt{x = x<<1} & $1.0$ & $0.0$ & $0.8$ & $0.0$\\ \cmidrule{2-7}
& \texttt{break} & \texttt{\textcolor{red}{sys.exit()}} & $0.0$ & $0.0$ & $0.1$ & $0.0$\\ \cmidrule{2-7}

\textit{Python}  & \texttt{for byte in encoder:} & \texttt{for bytes in encoder:} & $1.0$ & $0.0$ & 0.95 & $0.0$\\ \cmidrule{2-7}
 
& \texttt{encoded = "\textbackslash \textbackslash x"} & \texttt{encoded = ‘\textbackslash \textbackslash x'} & $1.0$ & $0.0$ & $0.87$ & $0.0$\\ \midrule 

\end{tabular}
\end{table}

%\todo[noline,size=\footnotesize]{\textbf{Rev 3.1}}\revision{
Our analysis highlights that n-gram-based metrics like ROUGE and BLEU, which are commonly used to assess code generation tasks, are not the best choice to evaluate offensive code. Indeed, the exact match and the edit distance are the most correlated to human evaluation for the assembly and Python code, respectively.
Assembly instructions are typically characterized by a fixed and concise structure in the form \texttt{OP DST, [SRC]}, which includes an \textit{opcode}, the destination of the operation, and (optionally) the source.
Therefore, increasing the \textit{n} value for these metrics (i.e., 3-4) leads to lower scores even for semantically correct snippets. 
The same goes for Python, which, although being a high-level and structurally complex language, when used for offensive purposes is characterized by concise snippets that handle numerical values and logical operations. 
%Contrarily, metrics such as the edit distance and the exact match do not take into account the code's length, but only its similarity to the reference. Therefore, the aforementioned examples receive a score equal to 1.
Table~\ref{tab:metrics_examples} shows a set of cherry-picked examples in both assembly and Python languages and their output similarity metric scores. We report the score for ROUGE-4, which is the least correlated metric with human evaluation for both languages, %and is representative of other n-gram-based metrics (e.g. BLEU-4), 
and edit distance and exact match, which are the most correlated for Python and assembly, as shown in \S{}~\ref{subsec:correlation_analysis}. The first and third rows for the assembly dataset present situations in which the predicted code matches the ground truth ($EM = 1$, $ED = 1$), yet the snippet is too short to be correctly assessed by ROUGE-4. The second row shows an example in which the reference and prediction are almost identical, but not semantically equivalent, therefore both ROUGE-4 and edit distance give a wrongly high score, while the exact match is correct.
Also for Python, the code does not contain 4-grams ($ROUGE$-$4 = 0$). The table shows two correct examples in which ground truth and predicted code differ by one or two characters but are equivalent, therefore the most accurate metric is the edit distance. The first row presents an example in which the prediction is completely inaccurate and all three similarity metrics correctly provide a low score.%}

Therefore, an important takeaway of our experiments is that the choice of metric depends on the complexity of the model-generated code. In the case of code with a more fixed structure and more limited set of instructions (as in the case of assembly), and, therefore, with less possibility of expressing different but semantically equivalent snippets, then the exact match is the best candidate. When the complexity of the generated code increases (as in the case of Python), then the edit distance metric is an appropriate choice for evaluating the code.
Moreover, when tasks are extremely difficult, i.e., when models fail to generate code (e.g., this is the common case of corpora not being large enough to train the models), the exact match is again the best metric. In the opposite case, i.e., when the models are very accurate in the code generation, then a metric that evaluates syntactic correctness (e.g., compilation accuracy) is recommended.

%\todo[noline,size=\footnotesize]{\textbf{Rev 1.1}}\revision{
Finally, we compared the results of our analysis with the results performed by previous studies in the generation of general (i.e., not offensive) Python code~\citep{DBLP:journals/corr/abs-2208-03133,DBLP:conf/profes/TakaichiHMKKKT22}. Since there is no further existing assembly dataset for code generation, we limited this analysis to the Python code.
ROUGE-L is found to be among the best-performing metrics for the assessment of the Python code on both the CoNaLa dataset~\citep{yin2018learning}, a dataset of questions posted on Stack Overflow with the posted solutions in Python, and the Card2code Hearthstone~\citep{DBLP:conf/acl/LingBGHKWS16}, a dataset dedicated to generating classes that are descriptions of the cards used in the Hearthstone game. A correlation analysis on the ReCa dataset~\citep{liu2020deep}, instead, showed the METEOR is the best metric to assess syntactically incorrect Python code generated from NL requirements.
Although these metrics act well also in the generation of offensive Python code, they do not result to have the highest correlation in our analysis. We attribute this result to the difference between generic code and offensive code.
Indeed, unlike regular code generation tasks that focus on logically complex functional code fragments, high-level exploit code contains a large number of arithmetic and logic operations, and bit-level slices (as in symmetric key cryptography) to encode the plain exploits into new, functionally equivalent ones, but more difficult to block by modern antivirus and intrusion detection systems.%} 

At the end of the day, despite output similarity metrics providing estimates close to the human evaluation, their ability to represent human assessment is highly affected by the specific code generation task, i.e., there is not a metric that is always suitable for the evaluation, regardless of data and complexity of the task.
Therefore, given that automatic code generation is an area that is likely to continue to attract interest from academia and industry, we believe there is a need for a solution that can automatically evaluate the semantic correctness of code generated by ML models.

%% file: threats.tex
\noindent
\textbf{AI-based code generators:} We performed our experiments employing two state-of-the-art NMT solutions, a Seq2Seq model and a pre-trained model such as CodeBERT. We are aware that considering only two models can be a limitation to this evaluation, yet our choice was guided by the popularity and the availability of mature open-source implementation of these technologies. 
Seq2Seq is still largely used as a baseline model in this line of research and remains among the most used architectures for code generation. CodeBERT, on the other hand, represents the state-of-the-art for several code-related tasks, such as code search and code documentation generation, and many other software engineering tasks~\citep{zhou2021assessing, DBLP:conf/msr/MashhadiH21, DBLP:conf/icse/AhmedD22,yu2022bashexplainer,zeng2022extensive}, including generation of offensive code~\citep{yang2023exploitgen,liguori2022can}.
%\todo[noline,size=\footnotesize]{\textbf{Rev 1.4}}\revision{
We acknowledge that there are emerging NMT models that are showing superior performance in different tasks, including code generation. However, the scope of this paper is not to improve the state-of-the-art performance in offensive code generation but to assess the ability of the metrics in estimating offensive code correctness. 
We believe that both Seq2Seq and CodeBERT fit well with the scope of the paper as they provide us with different and adequate numbers of code snippets to properly evaluate the metrics. Both assembly and Python code generated by the models are pretty balanced data (in terms of semantic correctness), which allows us to perform a fair evaluation of the metrics. Moreover, balanced data also enable the execution of different analyses, such as the analysis of whole wrong and whole correct data.%} 
Finally, we did not consider public AI code generators such as GitHub Copilot and OpenAI ChatGPT, since they impose restrictions on malicious uses~\citep{4}. Moreover, both attackers and defenders need to avoid leaking their techniques and tactics to their counterparts (``\textit{operations security}'', OPSEC). Thus, we consider the case of an attacker or defender that builds her own AI code generator, thus circumventing usage policies of public AI code generators. 
%We assess the feasibility of building an AI code generator using a deep learning model trained locally. We use an encoder-decoder architecture with attention mechanism, with a bi-directional LSTM as the encoder, which is a popular deep learning architecture for NLP. 

\noindent
\textbf{Dataset:} This work addresses the specific problem of the automatic generation of software exploits, focusing on the translation of NL intents into offensive code snippets in assembly and Python programming languages. 
%We considered several options for our evaluation. However, other datasets (e.g., JuICe \citep{DBL:conf/emnlp/AgasheIZ19}, HumanEval \citep{Chen2021EvaluatingLL}) contain relatively-small programs that are only commented on at a rather high level. For instance, JuICe includes programming assignments with only one statement to describe the entire program to be generated.
The datasets we used in our experiments fit perfectly with the scope of this work. Indeed, to the best of our knowledge, this is the only dataset used for code generation in the context of software security.
Moreover, we aimed to address complex and longer programming tasks (e.g., Python and assembly for processing binary-level data). For these tasks, NMT is still far from generating long and complex programs from just a single high-level description. The assembly and Python datasets considered in this work provide natural language descriptions both at the block and statement levels that are closer to the descriptions needed for more complex programming tasks.

\noindent
\textbf{Data size:} We acknowledge that the corpora used in our experiments may seem relatively small compared to other corpora available for different code-generation tasks. These corpora contain relatively-small programs that are described at a rather high-level (e.g., the JuICe dataset~\citep{agashe2019juice} includes programming assignments with only one statement to describe the entire program to be generated) or larger, potentially noisy, subsets of training examples obtained by mining the web (e.g., CoNaLa \textit{mined} dataset contains thousands of training examples mined directly from  StackOverflow~\citep{yin2018learning}).
The datasets used for our experiments, instead, are manually curated datasets containing high-quality and non-ambiguous descriptions of the code (not available in larger datasets for code generation) that help us to properly estimate whether the model's prediction is the correct translation of the NL intent.
Nevertheless, to mitigate the bias, we leverage an existing pre-trained model such as CodeBERT to compensate for the need for big data. Finally, it is worth noticing that, when we limit the comparison to manually annotated datasets, the dataset used in our experiments is way larger than the size of the CoNaLa \textit{annotated} dataset~\citep{yin2018learning}, which is the basis for state-of-the-art studies in NMT for Python code generation \citep{yin2019reranking,gemmell2020relevance}.

%Moreover, to the best of our knowledge, this is the first work to address the issue of correlating output similarity metrics to human judgment for low-level programming languages such as assembly. 

%We performed our evaluation of the quality of the generated code using non-code-oriented metrics. 
%We are aware that, recently, several metrics have been proposed to overcome the shortcomings of BLEU for code generation, such as CodeBLEU \citep{DBLP:journals/corr/abs-2009-10297} and RUBY \citep{DBLP:conf/iwpc/TranTNNN19}.
%To select a comprehensive set of output similarity metrics, we delved into the context of the evaluation of output code through automatic metrics, ranging from code generation (i.e., natural language to code), to code translation (i.e., programming language to different programming language), to code completion (i.e., programming language to the same programming language) tasks.

\noindent
\textbf{Output Similarity Metrics}:
We performed our evaluation by using a comprehensive and not trivial number of metrics used by previous work to assess the code generated by the NMT models. 
Nevertheless, we are aware that the set of metrics considered in our study does not include all the available metrics in the literature.
This is the case of the code-oriented metrics (i.e., metrics created ad-hoc for specific programming languages) proposed to overcome the shortcomings of \textit{BLEU} for code generation (e.g., \textit{CodeBLEU} \citep{DBLP:journals/corr/abs-2009-10297} and \textit{RUBY} \citep{DBLP:conf/iwpc/TranTNNN19}, which were designed to evaluate code written in Java and C\#).
However, these metrics rely on deeper program analysis (such as syntax and dataflow match), which requires that the code generated by the models is syntactically correct (i.e., compilable) and prevents the metrics from being language-agnostic. As matter of fact, there is no available implementation of such metrics for low-level programming languages such as assembly.
Therefore, we focused on the output similarity metrics commonly applied for the code evaluation, which can be easily tuned and used, regardless of the code programming language.

%% file: conclusion.tex
%In this practical experience paper, we assessed the ability of the output similarity metrics for the evaluation of the NMT models in the generation of code for software security applications.
In this work, we compared the results provided by the output similarity metrics with the human evaluation by assessing the performance of two different state-of-the-art models in the generation of offensive assembly and Python code snippets from natural language descriptions.

The results of our experiments provide actionable insights that can be used by future research to assess the NMT models in the software security field.
Although we pointed out what metrics can properly assess the model's predictions in different case studies, there is still a gap to fill between the automatic and the human evaluation. 
%\todo[noline,size=\footnotesize]{\textbf{Rev 1.3}}\revision{
Unfortunately, using a simple metric (however smart) is often not sufficient to provide accurate results. Therefore, to have significant improvements, we believe it is necessary to apply static and dynamic analysis techniques to automatically assess the semantic correctness of code generated by the models.%}

%Indeed, our experiments showed that the metrics commonly used for the evaluation of the models, such as \textit{BLEU-4}, provide results very different from the human evaluation and have a low correlation with the semantic correctness of the code. We also found that n-gram-based metrics provide results close to the human evaluation when the number of n-grams is low. 
%Finally, the correlation analysis showed that the \textit{exact match} and the \textit{edit distance} are the most correlated metrics with the semantic correctness of the assembly and Python code, respectively.